\documentclass[aps,prl,twocolumn,floatfix,amssymb]{revtex4} 
\usepackage{epsfig} 

\newcommand{\pe}[1]{#1_{\bot}} 
\newcommand{\pa}[1]{#1_{\|}}

\newcommand{\su}[2]{\frac{#1}{#2}}

\begin{document} 
 
%\draft 
%\twocolumn[\hsize\textwidth\columnwidth\hsize\csname@twocolumnfalse\endcsname 
 
\title{Transverse Fluctuations in the Driven Lattice Gas} 
 
\author{Sergio Caracciolo} 
\email{Sergio.Caracciolo@mi.infn.it} 
\affiliation{Universit\`a degli Studi di Milano -- Dip.~di Fisica and INFN, via Celoria 16, I-20133 Milano, and NEST-INFM, Italy} 
\author{Andrea Gambassi} 
\email{Andrea.Gambassi@sns.it} 
\affiliation{Scuola Normale Superiore and INFN -- Sez. di Pisa, 
Piazza dei Cavalieri 7, I-56100 Pisa, and INFN, Italy} 
\author{Massimiliano Gubinelli} 
\email{mgubi@cibs.sns.it} 
\affiliation{Dipartimento di Matematica Applicata and INFN -- Sez. 
 di Pisa, Universit\`a degli Studi di Pisa, I-56100 Pisa, Italy} 
\author{Andrea Pelissetto} 
\email{Andrea.Pelissetto@roma1.infn.it} 
\affiliation{Dipartimento di Fisica and INFN -- Sez. di Roma I, Universit\`a degli Studi di Roma ``La Sapienza'', I-00185 Roma, Italy}

\date{\today} 
%{\small 
%\begin{center} 
%{\bf \abstractname} 
%\end{center} 
\begin{abstract} 
We define a transverse correlation length 
suitable to discuss the finite-size-scaling behavior of an out-of-equilibrium 
lattice gas, whose 
correlation functions decay algebraically with the distance. 
By numerical simulations we verify that this definition has a 
good infinite-volume limit independent of the lattice geometry. 
We study the transverse fluctuations as they can select the correct
field-theoretical description. By means of a careful finite-size
scaling analysis, without tunable parameters, we show that they are Gaussian,
in agreement with the predictions of the model proposed by  Janssen,
Schmittmann, Leung, and Cardy.  
\end{abstract} 
 
%\pacs{} 
%\keywords{non-equilibrium, finite size scaling, strong anisotropy, non-Gibbsian state, Kawasaki dynamics} 
 
%] 
 
\maketitle 
 
%%%%%%%%%%%%%%%%%%%%%%%%%%%%%%%%%%%%%%%% 
%%%%%%%%%%%%%%%% CORPUS %%%%%%%%%%%%%%%% 
%%%%%%%%%%%%%%%%%%%%%%%%%%%%%%%%%%%%%%%% 
 
While the statistical mechanics of systems in thermal equilibrium is 
well established, no sound framework is available for nonequilibrium 
systems, although some interesting results have been 
recently obtained~\cite{Derrida}. 
At present, most of the research focuses on very specific models. 
Among them, the driven lattice gas (DLG), introduced by 
Katz, Lebowitz, and Spohn~\cite{KLS}, has attracted much attention 
since it is one of the simplest nontrivial models with a nonequilibrium 
steady state \cite{Zia}. The DLG is a generalization of the lattice gas. 
One considers a hypercubic lattice and for each site $x$ introduces an occupation 
variable $n_{x}$, which can be either zero (empty site) 
or one (occupied site). Then, one introduces an external field 
$E$ along a lattice direction and 
a generalization of the Kawasaki dynamics for the lattice gas with 
nearest-neighbor interactions. In practice, one randomly chooses a lattice 
link $\langle xy\rangle$, and, if $n_{x}\not= n_y$, proposes a particle 
jump which is accepted with Metropolis probabilities $w(\beta \Delta H + \beta E \ell)$, where $\ell = (1,0,-1)$ for jumps (along, transverse, opposite) to the 
field direction, $w(x) = \hbox{\rm min}\, (1,e^{-x})$, 
 and $\Delta H$ is the variation of the standard lattice-gas nearest-neighbor interaction 
\begin{equation} 
H = - 4\,\sum_{\langle xy\rangle} n_x n_y. 
\end{equation} 
As usual, the 
parameter $\beta$ plays the role of an inverse temperature. A nontrivial 
dynamics is obtained by considering periodic boundary conditions 
in the field direction. Indeed, in this case a particle current sets in, 
giving rise to a nonequilibrium stationary state that is non-Gibbsian. 
 
At half filling, the DLG undergoes a second-order phase transition. 
Indeed, at high temperatures the steady state is disordered while at low 
temperatures the system orders: the particles condense forming a strip 
parallel to the field direction. 
The two temperature regions are separated by a phase transition 
occuring at the critical value $\beta_c(E)$ depending on the field $E$. 
$\beta_c(E)$ converges to the Ising critical value $\beta_I$ for $E\to 0$, and, interestingly enough, increases with $E$, and 
saturates at a finite value $\beta_{c}(\infty)$ when $E$ diverges. 
Such a transition is different in nature from the order/disorder one 
occuring in the lattice gas. For instance, 
it is {\em strongly anisotropic}, i.e. fluctuations in 
density correlation functions behave in a 
qualitatively different way depending on whether one considers points 
that belong to lines that are parallel or orthogonal to the field $E$. 
 
Some time ago Janssen, Schmittmann, 
Leung, and Cardy \cite{Janssen86a} (JSLC) proposed a 
Langevin equation for the coarse-grained density field, 
which incorporates the main features of the DLG, i.e. 
a conserved dynamics and the anisotropy induced by the external 
field and should therefore describe the critical behavior of the 
DLG. By means of a standard Renormalization-Group (RG) analysis, 
critical exponents were exactly computed in generic 
dimension $d$ for $2<d<d_{c}=5$. 
This model predicts a strongly anisotropic 
behavior with correlations that increase with different exponents 
$\pa{\nu}$ and $\pe{\nu}$ 
in the directions parallel and orthogonal to the external field~\cite{Zia}. 
The corresponding anisotropy exponent $\Delta =(\pa{\nu}/\pe{\nu})-1$ 
can be exactly computed finding $\Delta = (8-d)/3$. Another remarkable 
property of the JSLC model is that transverse fluctuations become 
Gaussian in the critical limit. Extensive numerical simulations~\cite{china} 
have confirmed many predictions of the JSLC model, although 
some notable discrepancies still remain. 
The conclusions of JSLC have been recently questioned by 
Garrido, de los Santos, and Mu\~noz \cite{Garrido}. 
They argued that the DLG for $E=\infty$ is not described 
by the JSLC model but is rather in the same universality class 
as the Randomly Driven Lattice Gas (RDLG). 
This model has upper critical dimension $d_{c}=3$. The RG analysis done in 
Ref.~\cite{RDLG} leads to series expansions for the critical exponents in 
powers of $\epsilon\equiv d_{c}-d$. In particular, $\Delta = 1 + O(\epsilon^2)$, 
consistent with $\Delta\approx 1$ in $d=2$ (the case we will consider in our numerical simulations), quite different from the JSLC 
prediction $\Delta=2$. A second notable difference is that in the RDLG 
transverse critical fluctuations are not Gaussian. 
Apparently, numerical simulations of the DLG~\cite{AGMM} are also 
in agreement with the RDLG scenario and consistent with $\Delta\approx 1$ 
in two dimensions. 
 
In view of these contradictory recent results, 
new numerical investigations are necessary, 
in order to decide which of these two models really describes the DLG universality class. 
In this paper, we will focus on the transverse fluctuations for the 
very simple reason that in the JSLC model there is a very strong prediction: 
transverse correlation functions are Gaussian so that, 
beside critical exponents, one can also compute exactly 
the scaling functions of several observables. 
Therefore, it is possible to make stronger tests of the 
JSLC model. 
 
In order to test the JSLC predictions, we will investigate the 
Finite-Size Scaling (FSS) behavior of several observables. 
For an isotropic model on a lattice $L^d$, the FSS limit corresponds to 
$t\equiv 1-\beta/\beta_{c}\to 0$, $L\to\infty$, keeping 
$tL^{1/\nu}$ constant. 
This has to be modified for strongly anisotropic models~\cite{Binder} like 
the DLG. 
It has been argued that, for a geometry $\pa{L}\times \pe{L}^{d-1}$, 
one has to keep fixed 
both combinations $t\pa{L}^{1/\pa{\nu}}$ and 
$t\pe{L}^{1/\pe{\nu}}$, and therefore also the 
so-called {\em aspect ratio} $S_{\Delta}= \pa{L}^{1/(1+\Delta)}/\pe{L}$. 
Then, if an observable $\cal O$ diverges 
at criticality as $t^{-z_{\mathcal{O}}}$, in a finite lattice one has 
\begin{equation} 
{\cal O}(\beta;\pa{L},L) \approx L^{z_{\cal O}/\pe{\nu}} 
f_{{\cal O}}(t^{-\pe{\nu}} /L,S_{\Delta}), 
\label{anisFSS} 
\end{equation} 
where we have neglected subleading scaling corrections. To simplify the notation we write $L$ instead of $\pe{L}$ and, below, $\xi_L$ for the {\em transverse} correlation length. 
In many numerical studies Eq. (\ref{anisFSS}) has been 
tested for several observables, but this can be a very 
weak test since several parameters, 
$\beta_{c}, \pe{\nu}, z_{\mathcal{O}}$, and $\Delta$, 
must be tuned in order to fit the numerical data. 
A stronger FSS test can be performed if one uses 
a suitably defined correlation length. 
In this case, using a transverse (infinite volume) correlation length 
$\xi_{\infty}$, 
Eq.~(\ref{anisFSS}) may be written in the form 
\begin{equation} 
{\cal O}(\beta;\pa{L},L) \approx 
L^{z_{\cal O}/\pe{\nu}} \tilde{f}_{{\cal O}}(\xi_{\infty}(\beta)/L,S_{\Delta}). 
\label{eq:xiform} 
\end{equation} 
In this equation $\beta_{c}$ does not explicitly appear 
and $z_{\cal{O}}$ and $\pe{\nu}$ 
enter only through their ratio, cancelling when one looks directly 
at the correlation length. One can also eliminate this unknown by considering 
the ratio of the observables 
on two different lattices with transverse sizes $L$ and $\alpha L$. In this case 
we have 
\begin{equation} 
{{\cal O}(\beta;\alpha^{1+\Delta} \pa{L}, \alpha L)\over 
 {\cal O}(\beta; \pa{L}, L)} 
\approx 
F_{\cal{O}}\left({\xi(\beta;\pa{L}, L)\over L},S_{\Delta},\alpha\right), 
\label{eq:ourform} 
\end{equation} 
where only $\Delta$ has to be fixed {\em a priori}. 
 
In order to test Eqs.~(\ref{eq:xiform}) and (\ref{eq:ourform}), 
 one has to define a finite-size 
correlation length. In the DLG this is not obvious because correlation 
functions always decay algebraically at large distances~\cite{Grinstein}. 
A parallel correlation length was defined in Ref.~\cite{Valles8687}, 
but it suffers from many ambiguities (see the discussion in 
Ref. \cite{Zia}). The definition of a transverse 
correlation length is even more difficult, 
because of the presence of negative correlations at large 
distances~\cite{china,Zia}. 
 
In this paper we propose a new definition that 
generalizes the second-moment correlation length 
used in equilibrium systems. 
The basic observation is that the infinite-volume transverse the two-point \emph{wall-wall} 
correlation function decays exponentially, 
so that a transverse correlation 
length can be naturally defined in the thermodynamic limit. 
The extension of this definition to finite volumes requires some care because 
of the conserved dynamics, which makes the two-point function vanish at zero 
momentum. Here, we will use the results of Ref.~\cite{correlation_length}. 
Given the Fourier transform $\widetilde{G}({q})$ of the two-point correlation function $\langle n_{{x}}n_{{0}}\rangle$, 
we focus on the transverse correlation 
$\pe{\widetilde{G}}(q) \equiv \widetilde{G}(\left\{\pa{q}=0,\pe{q}=q\right\})$ and 
define 
\begin{equation} 
\label{defxi} 
 {\xi}_{ij} \equiv 
 \sqrt{\su{1}{\hat{q}_{j}^{2} - \hat{q}_{i}^{2}} 
 \left( \su{\pe{\widetilde{G}}(q_{i})}{ \pe{\widetilde{G}}(q_{j})} -1\right) }, 
\end{equation} 
where $\hat{q}_{n}=2\sin{(\pi n/L)}$\ is the lattice momentum. 
If the infinite-volume transverse wall-wall correlation function 
decays exponentially, $\pe{\widetilde{G}}(q)$ has a regular expansion 
in powers of $q^2$ and 
\begin{equation}
        \pe{\widetilde{G}}^{-1}(q)\approx \chi_{\infty}^{-1} [ 1 + b^{2}\, q^{2}+ O(q^{4})]\; ,\label{Gexp}
\end{equation}
where the coefficient $b$ of $q^2$ naturally defines a transvers correlation length $\xi_{\infty}$.
We expect Eq.~(\ref{Gexp}) to hold also in a finite box. 
Then, starting from Eq.~(\ref{defxi}), it is easy to show that 
${\xi}_{ij}$ converges to $\xi_{\infty}$ as $L\to \infty$, justifying our 
definition of finite-volume correlation length. 
In the subsequent analysis we consider $\xi_{13}$ as the 
finite-volume (transverse) correlation length 
$\xi_L(T) \equiv \xi(T;S_\Delta^{1+\Delta} L^{1 + \Delta}, L)$. 
As in previous studies, 
we also define a finite-volume transverse susceptibility as $\chi_{L} = \pe{\widetilde{G}}(2\pi/L)$. 
 
As we already said, in the JSLC model transverse fluctuations are Gaussian. 
This allows us to compute the scaling function appearing 
in Eq. (\ref{eq:ourform}) for $\xi_L$ and $\chi_L$. If $\pe{\widetilde{G}}(q)$ is Gaussian we have 
\begin{equation} 
F_{\xi}(z,S_2,\alpha) = \left[ 1- \left(1 -\alpha^{-2}\right)(2\pi)^2 z^2 
 \right]^{-1/2}, 
\label{eq:theoretical_pred} 
\end{equation} 
with $z\equiv \xi_{L}/L$ and 
$F_\chi(z,S_2,\alpha) = F_\xi(z,S_2,\alpha)^2$. We can also 
compute $\tilde{f}_\xi(x,S_2)$, see Eq.~(\ref{eq:xiform}), obtaining 
\begin{equation} 
\tilde{f}_\xi(x,S_2) = \left({ 4\pi^2} + {1/ x^2}\right)^{-1/2}. 
\label{eq:theoretical_pred2} 
\end{equation} 
\begin{figure}[!ht] 
 \begin{tabular}{c} 
 {\epsfig{file=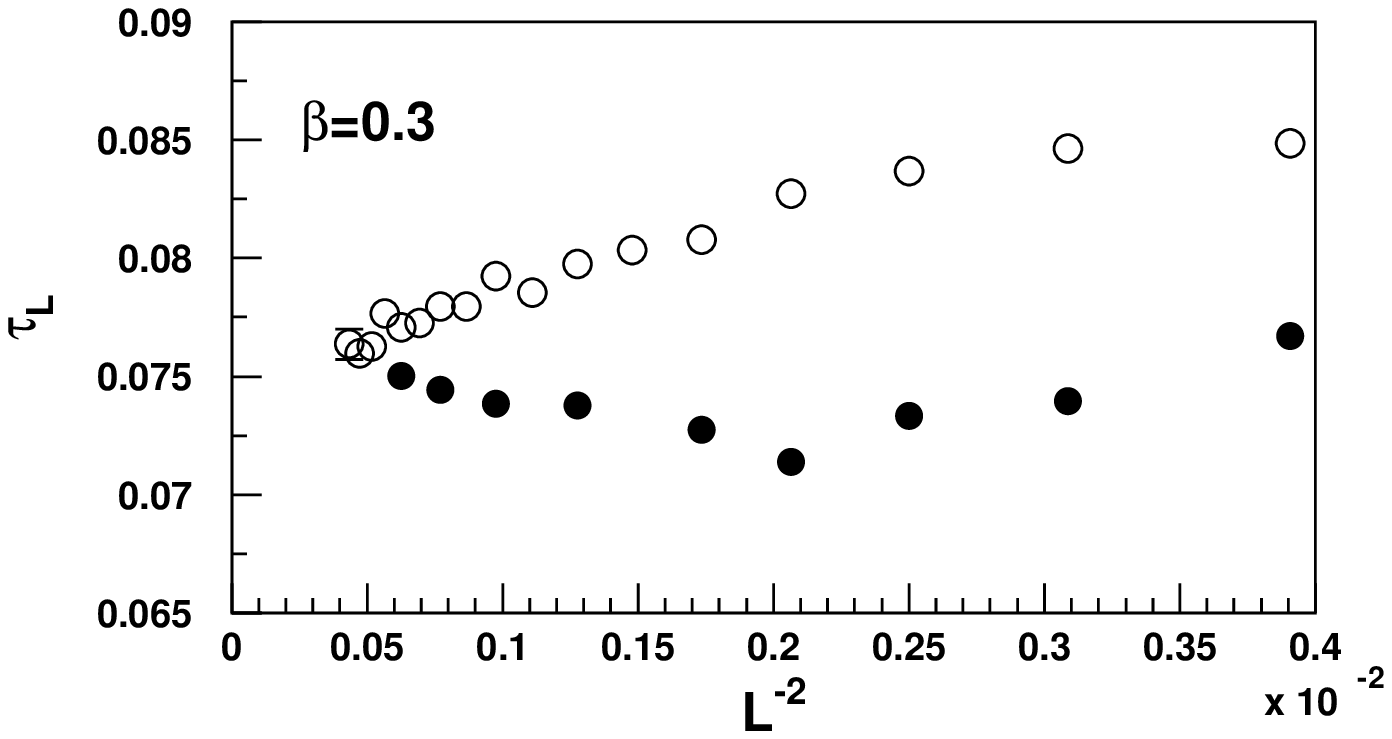,width=0.45\textwidth}}\\ 
 {\epsfig{file=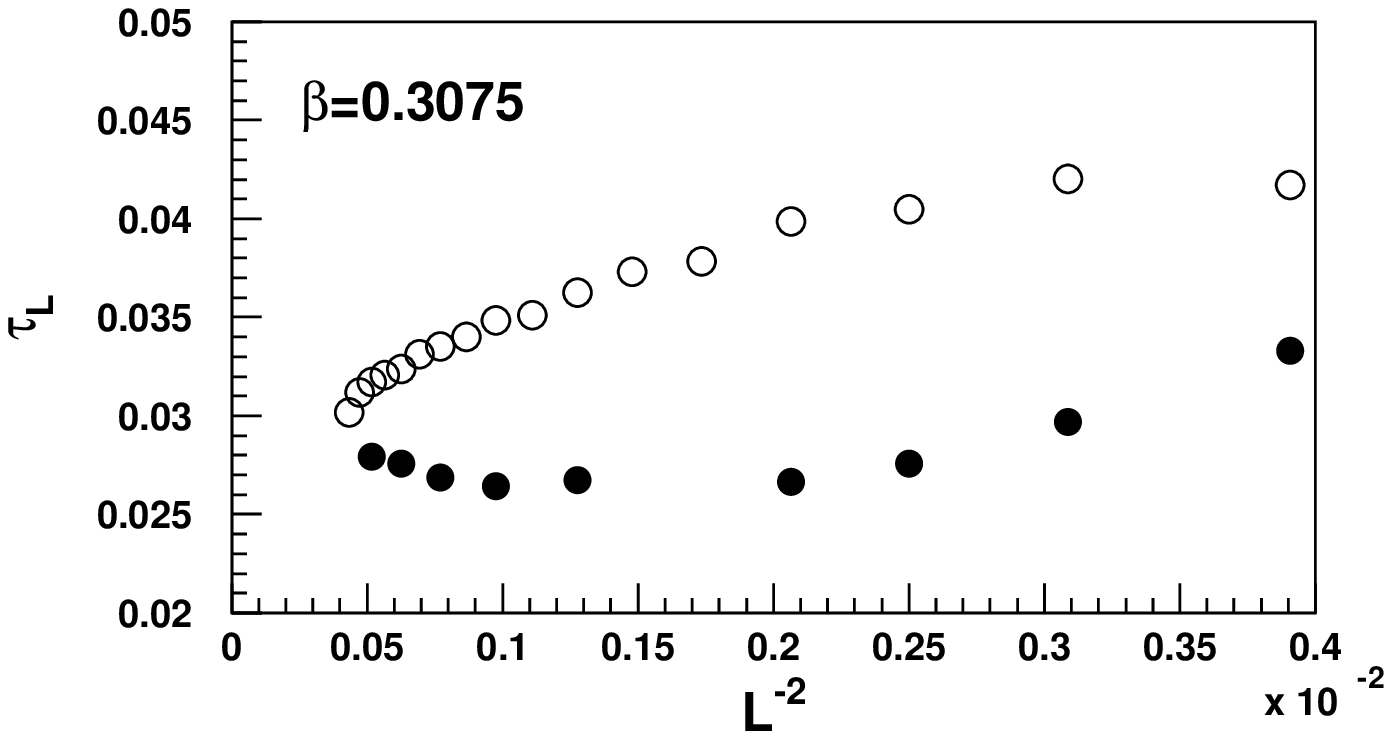,width=0.45\textwidth}}\\ 
 {\epsfig{file=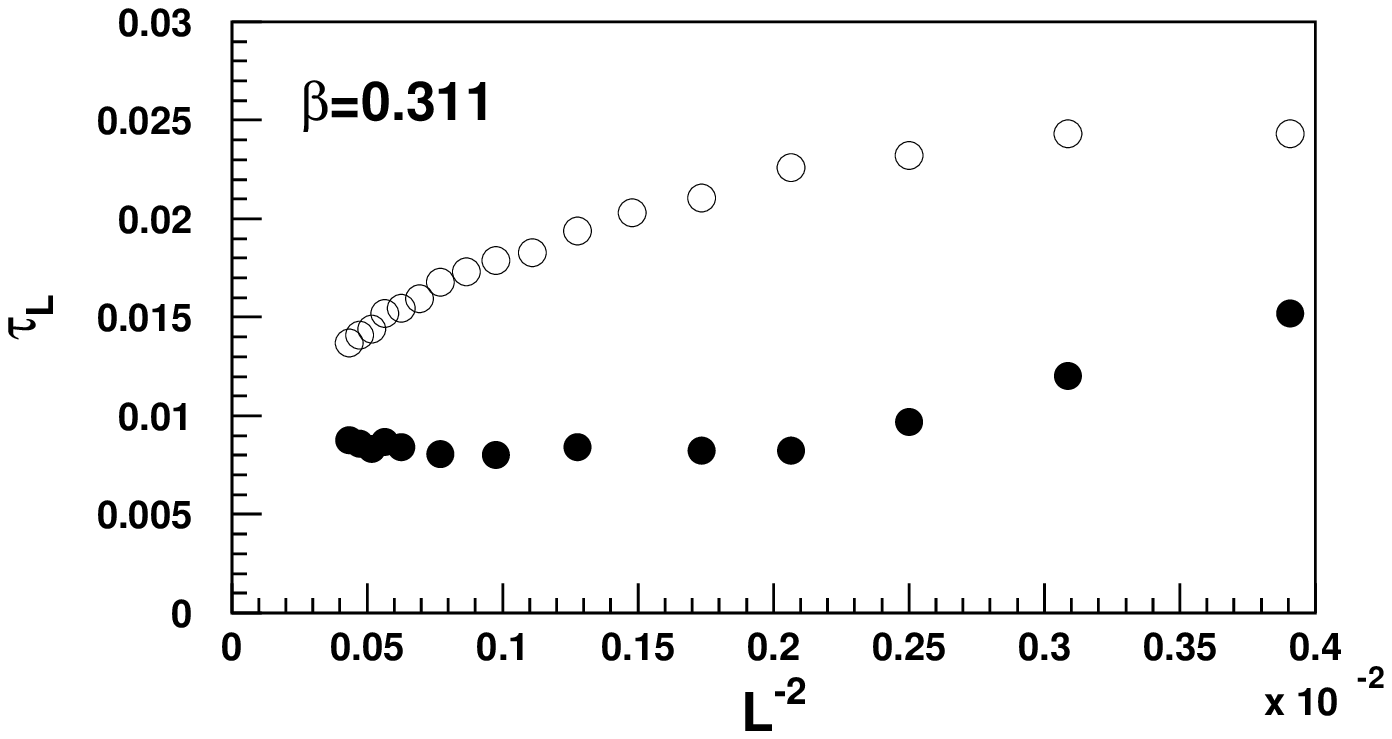,width=0.45\textwidth}}\\ 
 {\epsfig{file=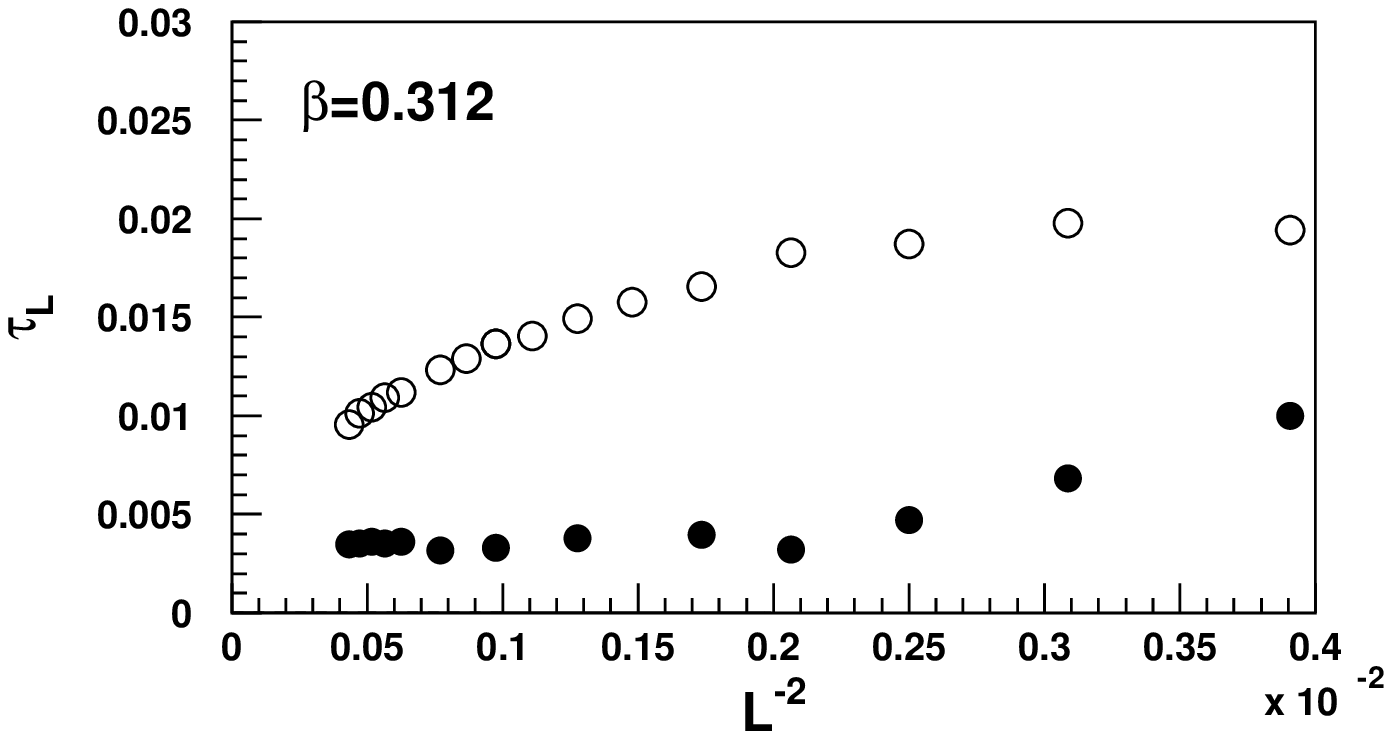,width=0.45\textwidth}}\\ 
 \end{tabular} 
\caption{$\tau_L$ for different 
inverse temperatures $\beta$. Filled (respectively empty) 
points refer to geometries with aspect ratio $S_{2}$ 
(respectively $S_{1}$) fixed. Here $S_2 \approx 0.200$, $S_1 \approx 0.106$. 
Errors are smaller than the size of the points. 
} 
 \label{fig:tauelle1} 
\end{figure} 
In this paper we want to make a high-precision test of 
the theoretical predictions of the JSLC model, 
Eqs.~(\ref{eq:theoretical_pred}) and 
(\ref{eq:theoretical_pred2}). We work in two dimensions 
at infinite driving field. 
Since we wish to test the JSLC predictions we fix $\Delta=2$ 
and consider lattice sizes with $S_2 \approx 0.200$. 
The largest lattice corresponds to $\pa{L}= 884$, $L = 48$. 
For each lattice size, we compute $\chi_L$ and $\xi_L$ 
for several values of $\beta$ 
lying between $0.28$ and 0.312. 
 
As a preliminary test, we verify that our definition of $\xi_{L}$ has a good 
thermodynamic limit. For this purpose, 
we introduce the following quantity (the reason for this definition 
will be presented below) 
\begin{equation} 
 \label{eq:definition_of_tauelle} 
 \tau_L(\beta) \equiv \xi_L^{-2}(\beta) - 4 \pi^2 L^{-2}. 
\end{equation} 
In Fig.~\ref{fig:tauelle1} 
we plot $\tau_{L}(\beta)$ versus $1/L^{2}$ at 
several inverse temperatures $\beta$. 
For each $\beta$, $\tau_L(\beta)$ converges to a finite 
constant, showing that our definition has a finite 
infinite-volume limit. Moreover, the same result is obtained 
by using sequences of lattices with $S_2$ or $S_1$ fixed: 
the result does not depend on the way in which $\pa{L}$ 
and $L$ go to infinity. As expected, 
when the temperature approaches the critical value, it is necessary to 
use larger and larger lattices to see the convergence to the infinite-volume 
limit. At $\beta$ fixed we expect the convergence to become eventually 
exponential in $L$. However, for lattices with $S_2$ fixed 
we observed an intermediate region of values of $L$ in which $\tau_L$ is 
apparently constant. Such a region widens as $\beta$ approaches the 
critical point and is therefore in excellent agreement with the relation 
\begin{equation} 
 \xi_{\infty}^{-2}(\beta) \approx \tau_L(\beta) = \xi_L^{-2}(\beta) - 4 \pi^2 L^{-2} 
\label{Eq11} 
\end{equation} 
in the FSS limit $L\to \infty$, $\beta\to \beta_c$, neglecting subleading corrections 
that will depend in general on the chosen value of $S_2$ and are particularly 
small for our choice $S_2 \approx 0.200$. 
Eq.~(\ref{Eq11}) immediately gives Eq.~(\ref{eq:theoretical_pred2}). 
Thus, the results presented on Fig.~\ref{fig:tauelle1} 
are perfectly consistent with $\Delta=2$ and the JSLC prediction 
for $\tilde{f}_\xi(x,S_2)$. 
 
\begin{figure}[h!] 
 \centering 
\epsfig{width=0.95\linewidth,file=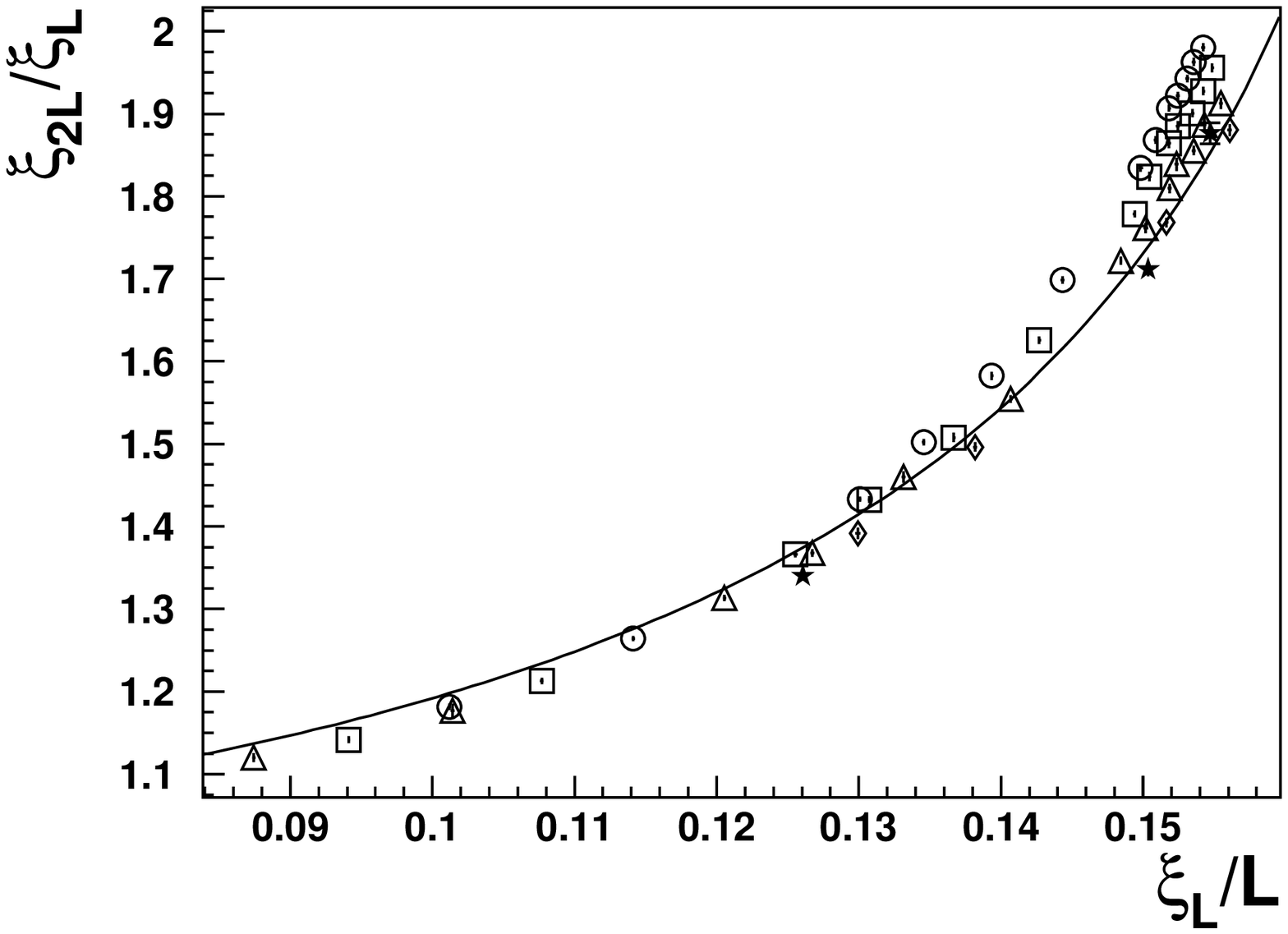} 
\caption{FSS curve for the transverse correlation length 
at fixed $S_{2}\approx 0.200$. 
Different symbols correspond to different lattice sizes: 
$L=$ $16(\circ )$, $18(\square )$, $20(\triangle )$, $22(\lozenge )$, 
 $24(\bigstar)$. 
The solid curve is the function $F_{\xi}(z,S_2,2)$ 
defined in Eq.~(\ref{eq:theoretical_pred}).} 
 \label{fig:xi} 
%\end{figure} 
 
%\begin{figure}[t] 
 \centering 
\epsfig{width=0.95\linewidth,file=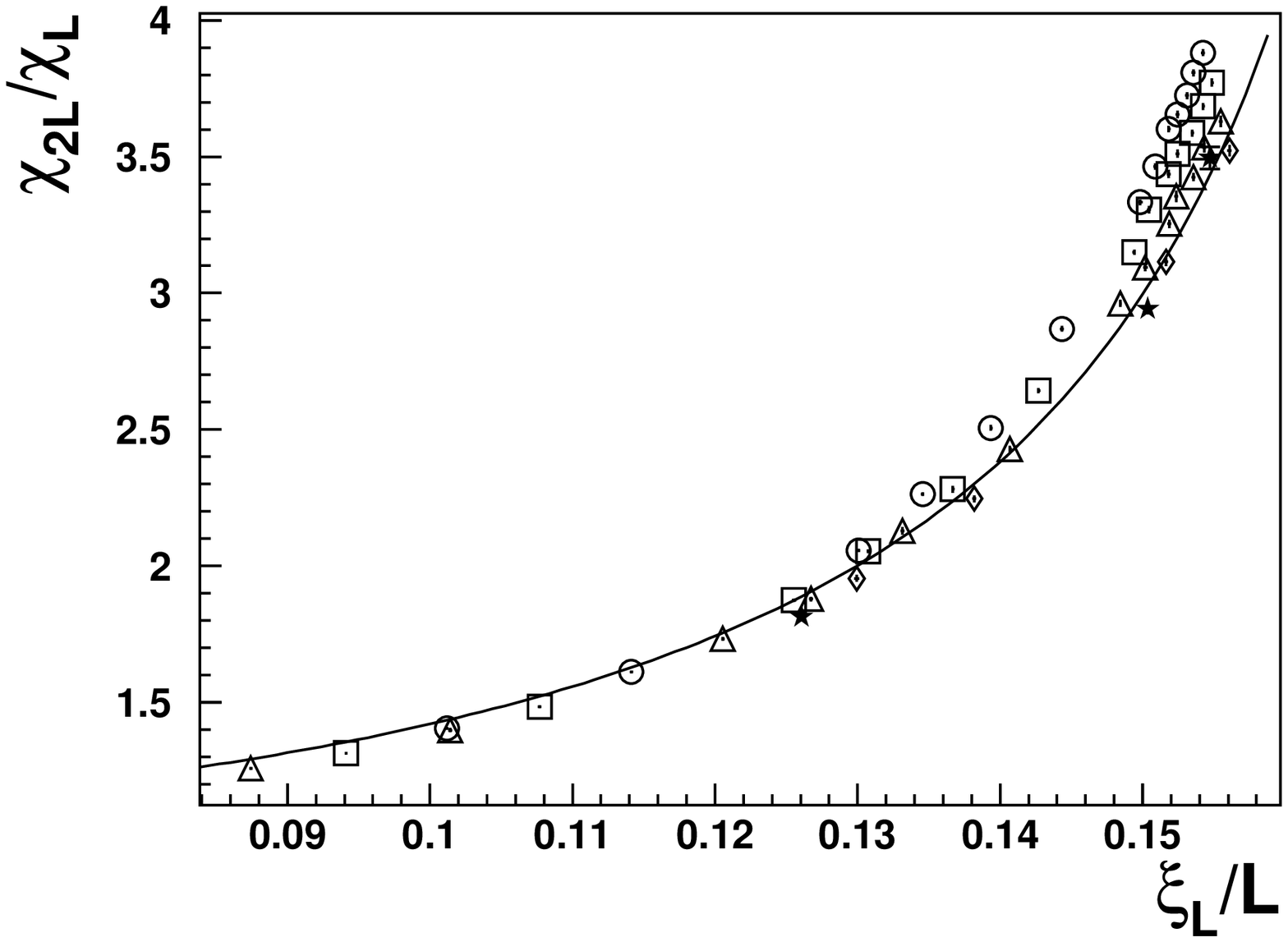} 
\caption{FSS curve for the transverse susceptibility 
at fixed $S_{2}\approx 0.200$. We use the same symbols of Fig. 
\protect\ref{fig:xi}. 
The solid curve is the function 
$F_{\chi}(z,S_2,2) = F_{\xi}(z,S_2,2)^2$, where 
$F_{\xi}(z,S_2,2)$ is 
defined in Eq.~(\ref{eq:theoretical_pred}).} 
 \label{fig:chi} 
\end{figure}

We want now to make a precise test of Eq.~(\ref{eq:theoretical_pred}). 
In Fig.~\ref{fig:xi} we report the results of our simulations for the ratio 
$\xi_{2L}/\xi_{L}$ as a function of $\xi_{L}/L$ for 
lattice sizes with fixed $S_{2}$, and we compare them with 
Eq.~(\ref{eq:theoretical_pred}). 
We stress that the theoretical curve is not a fit to the 
data: there is no free parameter to be chosen! 
Even if the agreement is not perfect, we notice that the points 
closer to the theoretical curve correspond to larger lattices. 
 
Using the universal function $F_\xi(z,S_2,2)$, we can extrapolate our data to 
infinite volume using the general strategy of Ref.~\cite{Caracciolo}. 
Correspondingly, we obtain 
$ \beta_{c} = 0.31256(9)$ 
and verify that, for small $t$, 
$ \xi_{\infty} \sim t^{-1/2} $ 
as predicted by JSLC. 
 
We can perform the same test for the susceptibility. 
In Fig.~\ref{fig:chi} we report our numerical 
results for $\chi_L$ together with the theoretical JSLC prediction. 
We observe a good agreement between theory and Monte Carlo results. 
 
We have also measured the transverse Binder cumulant $g_L$, 
see Ref.~\cite{Zia} for the definition. 
At the critical point we observe that $g_L(\beta_c) \sim L^{-0.2}$ for $L\to \infty$. Thus, the 
Binder cumulant vanishes at the critical point, 
again in agreement with the idea that transverse fluctuations 
are Gaussian. 
 
Further details of our analysis are presented elsewhere~\cite{noi}. 
 
In conclusion, we have shown 
by means of a new FSS analysis and new Monte Carlo data 
that the critical behavior of transverse fluctuations in the DLG model is Gaussian. 
Indeed, the numerical data are in perfect agreement with 
Gaussian FSS functions. Note that our results are stronger than those 
presented in previous analyses. First, we check not only critical exponents 
but a full scaling function; second, in our comparisons 
there are no free parameters that can be tuned. 
Our results are in perfect agreement with the predictions of the JSLC model. 
 
We thank Miguel Mu\~noz  for useful discussions and we acknowledge the
support of the INFM trough the Advanced Parallel Computing Project at CINECA.

\end{document}